\documentclass[journal,10pt,twocolumn]{IEEEtran}  

\usepackage{cite}
\usepackage{amsmath,amssymb,amsfonts}
\usepackage{algorithmic}
\usepackage{graphicx}
\usepackage[dvipsnames]{xcolor}
\usepackage{booktabs,multirow,array,colortbl}
\usepackage[normalem]{ulem}         
\usepackage{newtxtext,newtxmath}
\usepackage{dblfloatfix}
\usepackage{url}

\DeclareMathOperator*{\argmin}{arg\,min}

\begin{document}
\title{Visual Feedback of Pattern Separability Improves Myoelectric Decoding Performance of Upper Limb Prostheses}
\author{Ruichen Yang, Gy\"orgy M. L\'evay, Christopher L. Hunt, D\'aniel Czeiner, \\Megan C. Hodgson, Damini Agarwal, Rahul R. Kaliki, and Nitish V. Thakor
\thanks{This work was funded under the National Institutes of Health Grant No. U44NS119842. \textit{(Ruichen Yang and Gy\"orgy M. L\'evay are co-first authors.) (Corresponding author: Ruichen Yang.)}}
\thanks{R. Yang and Dr. N. Thakor are with the Department of Electrical Engineering, The Johns Hopkins University, Baltimore, 21218 USA}
\thanks{Dr. C. Hunt, Gy. L{\'e}vay, D. Czeiner, M. Hodgson, D. Agarwal, and Dr. R. Kaliki are with Infinite Biomedical Technologies, LLC., Baltimore, MD, 21218 USA.}}

\maketitle

\begin{abstract}
State-of-the-art upper limb myoelectric prostheses often use pattern recognition (PR) control systems that translate electromyography (EMG) signals into desired movements. As prosthesis movement complexity increases, users often struggle to produce sufficiently distinct EMG patterns for reliable classification. Existing training typically involves heuristic, trial-and-error user adjustments to static decoder boundaries. \textit{Goal}: We introduce the Reviewer, a 3D visual interface projecting EMG signals directly into the decoder’s classification space, providing intuitive, real-time insight into PR algorithm behavior. This structured feedback reduces cognitive load and fosters mutual, data-driven adaptation between user-generated EMG patterns and decoder boundaries. \textit{Methods}: A 10-session study with 12 able-bodied participants compared PR performance after motor-based training and updating using the Reviewer versus conventional virtual arm visualization. Performance was assessed using a Fitts law task that involved the aperture of the cursor and the control of orientation. \textit{Results}: Participants trained with the \textit{Reviewer} achieved higher completion rates, reduced overshoot, and improved path efficiency and throughput compared to the standard visualization group. \textit{Significance}: The \textit{Reviewer} The Reviewer introduces decoder-informed motor training, facilitating immediate and consistent PR-based myoelectric control improvements. By iteratively refining control through real-time feedback, this approach reduces reliance on trial-and-error recalibration, enabling a more adaptive, self-correcting training framework. \textit{Conclusion}: The 3D visual feedback significantly improves PR control in novice operators through structured training, enabling feedback-driven adaptation and reducing reliance on extensive heuristic adjustments.

\end{abstract}


\section{Introduction}
\label{sec:introduction}
In the United States, limb loss resulting from trauma affects over 1.6 million individuals\cite{ziegler2008estimating}. Upper limb loss (ULL) accounts for approximately 20 percent of total amputations\cite{avalere2024}. These injuries significantly hinder individuals from performing basic functional tasks, like grasping and holding objects, thereby profoundly impacting their day-to-day activities\cite{darter2018factors}. One of the leading solutions to address this challenge involves the use of multi-articulated prosthetic devices, which aim to replicate the movement and function of the missing anatomical limb\cite{geethanjali2016myoelectric}.

\begin{figure}[!t]
\centerline{\includegraphics[width=0.9\columnwidth]{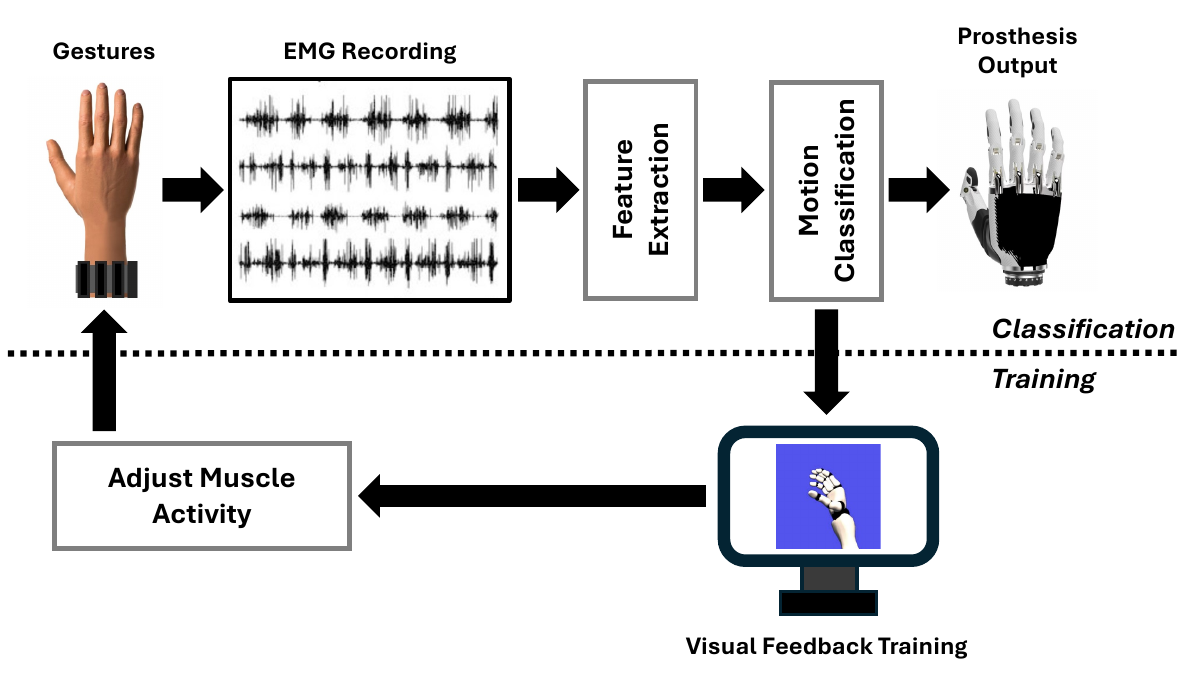}}
\caption{Motor-based training for PR-based prostheses. The process is structured into two parts: calibration and exploration. In calibration, users are prompted to perform muscle contractions as EMG signals are recorded. Features are extracted from these EMG signals and are used to train a PR-based motion classifier. During exploration, users observe classification output through modalities like a virtual arm and assess control performance. During this period, users can improve the quality of their control by: 1) adjusting muscle activity based on feedback; or 2) returning to the calibration phase to re-initialize the motion classifier with newly recorded EMG.}
\label{Demo}
\vspace{-5mm}
\end{figure}

Myoelectric control, which captures EMG signals generated during muscle contractions, is a widely used method for operating multi-articulated prosthetic devices. Users perform predefined muscle contractions with their residual limb while surface electrodes record the corresponding EMG signals. These signals are then mapped to prosthesis movements (e.g., hand open/close, wrist pronation/supination, elbow bend/extend) using various control techniques. Pattern recognition (PR) is one such technique, employing machine learning algorithms to classify EMG signals into distinct groups, each assigned to a specific movement\cite{parajuli2019real}. Popular PR methods for myoelectric prosthesis control include Support Vector Machines (SVMs), Artificial Neural Networks (ANNs), and Linear Discriminant Analysis (LDA)\cite{toledo2019support, alkan2012identification, hudgins1993new, phinyomark2012feature}.

 Under ideal conditions, PR-based myoelectric control offers the ability to control multiple degrees of freedom in a seamless manner\cite{kuiken2016comparison}.
However, experimental robustness does not necessarily equate to practical functionality. For novice users, there is often a steep learning curve to attain control proficiency\cite{bouwsema2014changes}. As the complexity and quantity of gestures employed in PR systems expand, the differentiation between each pattern becomes less discernible, leading to system confusion\cite{scheme2011electromyogram, bunderson2012quantification}.

To overcome this limitation, users typically undergo motor adaptation–based training designed to improve myoelectric control accuracy and robustness\cite{kristoffersen2019effect, bunderson2012quantification, powell2013training, he2015user}. Existing training methods primarily guide users to adapt their EMG patterns to the decoder’s static feature space by either increasing \textit{pattern separability} (i.e., maximizing interclass distances) or improving \textit{repeatability} (i.e., reducing within-class variability) \cite{bunderson2012quantification}. Commonly used protocols include motor imagery techniques, EMG-based training games incorporating proportional and derivative control, and 2D virtual-arm interfaces displaying user movements in real time\cite{powell2013user, dyson2020learning, davoodi2011real, van2016learning, winslow2018mobile} (\textcolor{RoyalBlue}{Fig. \ref{Demo}}). However, these approaches predominantly emphasize repeatability, offering limited feedback regarding pattern separability or the decoder’s decision boundaries \cite{Ottobock}. Recent studies have explored decoder-informed visual feedback to enhance myoelectric training. For instance, Montalivet et al. \cite{de2020guiding} employed a short-term, 2D "pattern similarity" display to correct specific problematic gestures, demonstrating improved real-time classification. Nawfel et al. \cite{nawfel2022influence} compared multiple visual feedback modalities (e.g., repeatability- and separability-based plots) within single-session protocols to evaluate their effects on training data quality. However, both approaches rely primarily on user adaptation to a static decoder, neglecting potential long-term benefits derived from mutual user-decoder adaptation. Consequently, these methods may be insufficient for addressing intrinsic EMG variability and signal drift over extended periods.

To address this gap, we propose a longitudinal 3D visual feedback system explicitly designed to facilitate continuous user-decoder co-adaptation. Unlike existing methods focusing primarily on contraction repeatability, our approach provides intuitive, real-time feedback on both EMG pattern separability and repeatability. Compared to traditional 2D or non-spatial visualizations, the proposed 3D feedback significantly reduces cognitive load by clearly representing spatial relationships among EMG patterns, thereby minimizing uncertainty and subjective interpretation. Additionally, our system incorporates an adaptive recalibration strategy in which both users and decoders iteratively refine their strategies—users adjust EMG input by observing the separability of muscle patterns in the classification-space, while the decoder updates classification boundaries via online calibration. This bidirectional adaptation is intended to accommodate natural variability in muscle activity, enhancing both short-term accuracy and long-term robustness of myoelectric control.

To evaluate our approach, we conducted a comparative longitudinal study involving an Experimental Group trained with our adaptive 3D visualization system and a Control Group trained using a conventional virtual-arm interface. By assessing changes in virtual task performance across multiple sessions, we tested the hypothesis that mutual adaptation would yield sustained and superior improvements compared to traditional user-centric methods.
Ultimately, we aim to demonstrate that adaptive, decoder-informed 3D visual feedback substantially promotes user-decoder co-adaptation, leading to improved task performance, increased decoder robustness, and enhanced overall control stability—outcomes unattainable by relying solely on user adaptation.

\begin{figure}[!t]
\centerline{\includegraphics[width=\columnwidth]{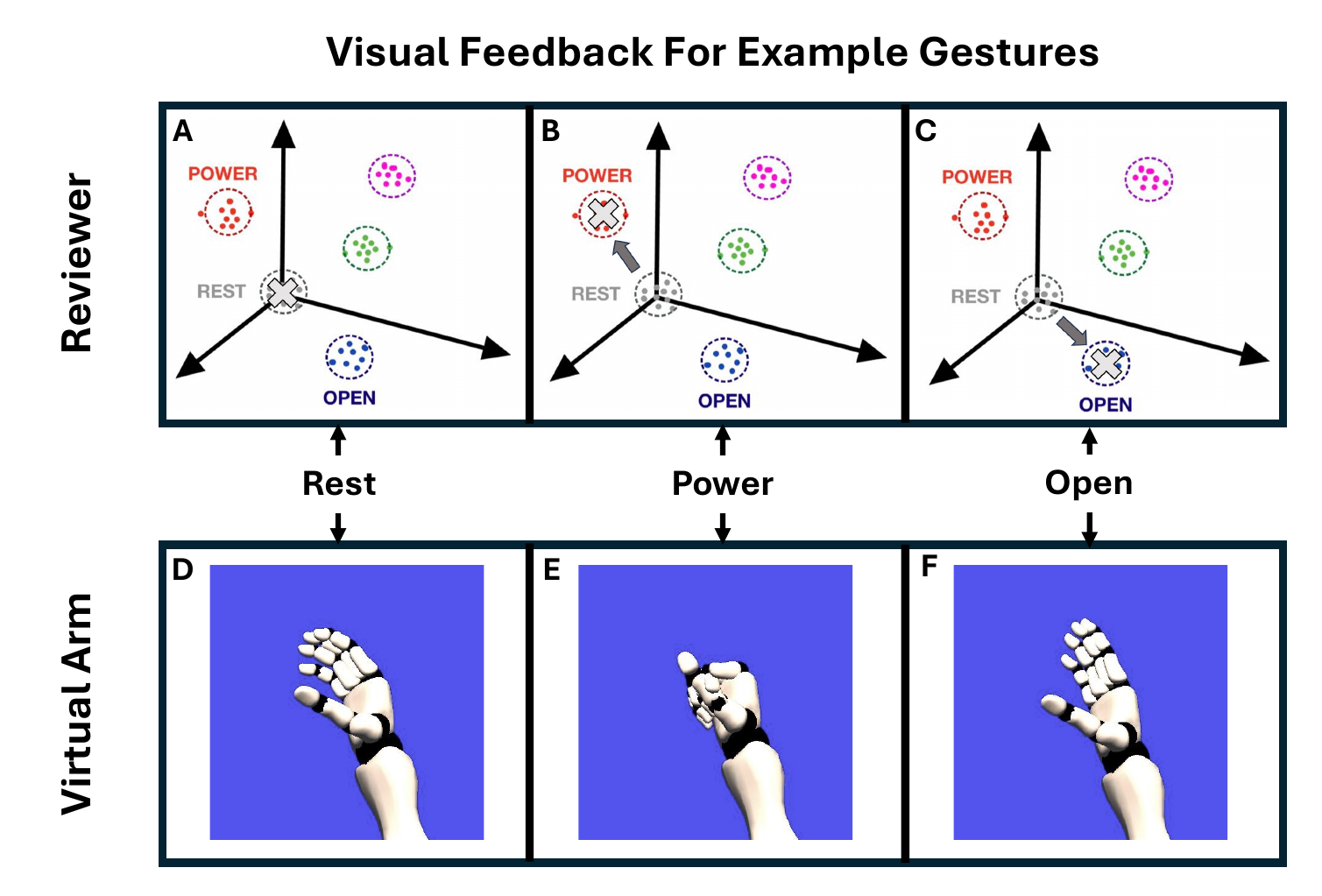}}
\caption{Representation of visualization feedback when example gestures are performed. (A-C) The \textit{Reviewer}: Using the PR-based control algorithm, EMG patterns for different gestures are clustered based on features extracted during the calibration phase. The resting position is marked as the initial point (0, 0, 0) in the feature space. Guided by patient data, the system optimizes the distribution of entries to maximize interclass variance and minimize within-class variance. After calibration, when the user performs a gesture, the white cursor moves to the corresponding location in the feature space as classified by the PR system. Subjects in the Experimental Group can access this system during the exploration phase and rotate the space for a comprehensive view. (D-F) Virtual Arm: When a gesture is performed, this system shows real-time movements of the arm as classified by the PR system. The Control Group accesses the virtual arm during the exploration phase.}
\label{EG_vs_CG}
\end{figure}

\section{Methods}
\label{sec: methods}
This study was conducted in accordance with protocol IRB00256686 approved by the Johns Hopkins University School of Medicine Institutional Review Board on Aug 23, 2021. Twelve able-bodied participants, 6 males and 6 females, were recruited to take part in a ten-session longitudinal experiment. Participants varied in age from 18 to 22. No participants had previous experience with PR-based myoelectric control. The experimental protocol undergone is first described in \cite{levay2024pattern}; however, its details are repeated below for completeness.

\subsection{Equipment}
Participant EMG signals were recorded using the Myoband, a wireless armband consisting of 8 channels of differential surface electrodes (Thalmic Labs, Ontario, Canada). The Myoband was positioned on the subject's dominant arm, over the cross-sectional area with the highest muscle mass. Additionally, the participant donned a bypass prosthesis socket consisting of a Bebionic Small multi-articulated hand (Ottobock GmbH, Duderstadt, Germany) and a MC Standard Wrist Rotator (Fillauer Motion Control, Salt Lake City, UT). This socket was securely attached to the dominant arm and was meant to replicate the weight-bearing experience of a prosthesis user with transradial limb loss.

\subsection{Pattern Recognition System}
\label{subsec:patrec}

\begin{figure}[!t]
\centerline{\includegraphics[width=\columnwidth]{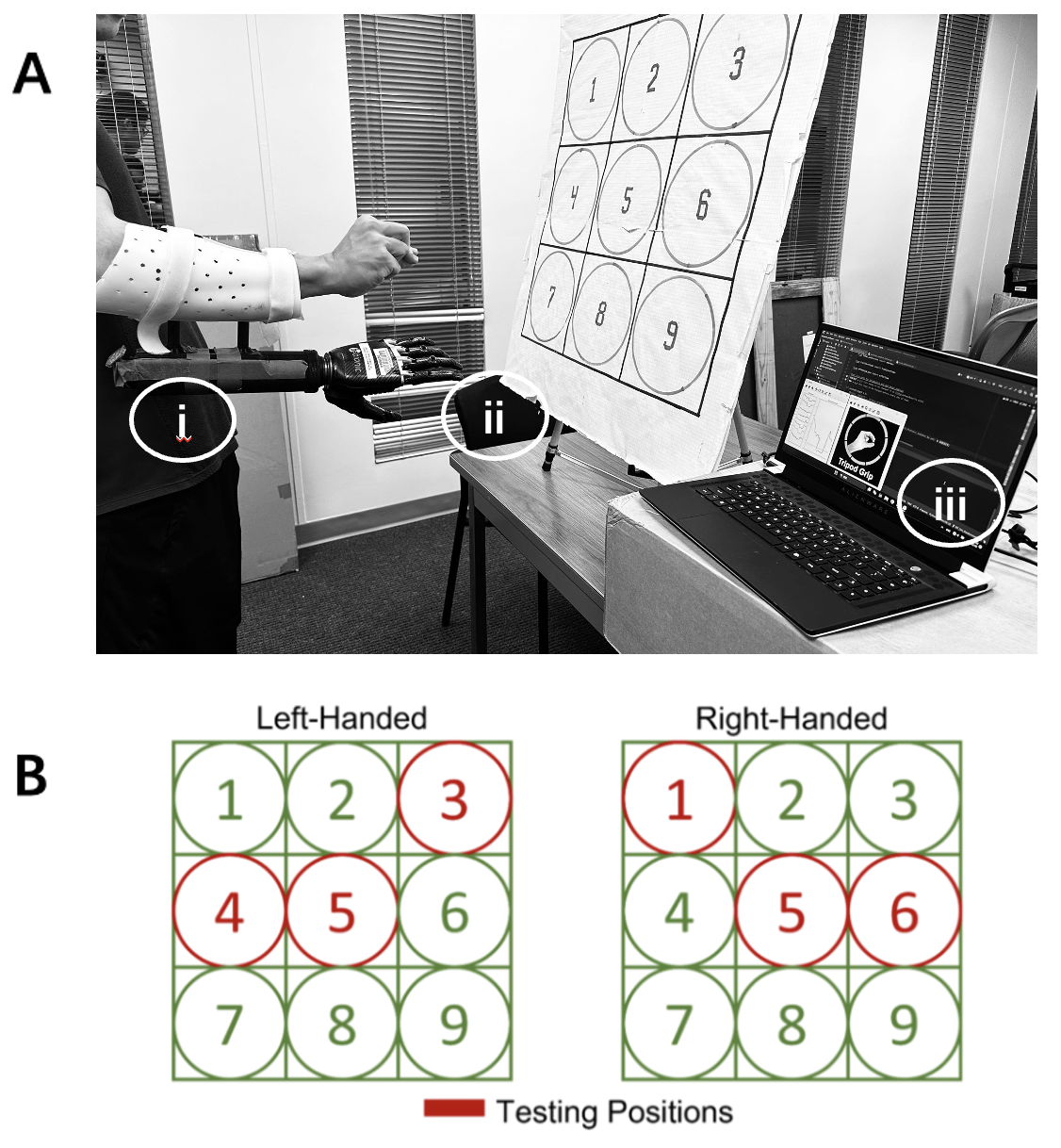}}
\caption{Experimental setup. (A) Equipment: i) A bebionic small multi-articulated hand was attached to each subject to simulate the weight-bearing experience of individuals with ULL. ii) A numbered board guided user movements, while iii) a display screen on the user's right showed gesture icons during calibration to prompt corresponding movements. In the exploration phase, the Experimental Group used a 3D visual system, while the Control Group used a virtual arm. (B) The FLT was administered during the testing phase, with subjects instructed to keep their dominant limb's shoulder aligned with Location 5, as indicated by the board. During testing, subjects placed the multi-articulated hand of the prosthesis in one of three testing locations. Left-handed individuals rotated between Locations 3, 4, and 5, while right-handed individuals alternated between Locations 1, 5, and 6.}
\label{setup}
\end{figure}

Features were extracted from incoming EMG signals with a 200 ms sliding window with 50 ms of overlap. For each channel of EMG, we extracted both time- and frequency-domain features \cite{englehart1999classification}. Movement prediction was then moderated using the \textit{Spatial Classifier}.

Given a calibration set of feature-extracted EMG samples $X = \left\{ X_0, X_1, ..., X_k \right\}$ where $X_i = \left\{ x_{i,0}, x_{i,1}, ..., x_{i,n} \right\}$ represents $n$ feature-extracted EMG datapoints $x_{i,j} \in \mathbb{R}^d$ collected during the $i$th of $k$ calibrated movements, the Spatial Classifier first employs LDA to project the feature-extracted EMG data into a lower-dimensional subspace\cite{englehart2003robust}. Within the subspace, the Spatial Classifier then defines a set of axes $L={L_0,...,L_{k-1}}$ where $L_i$ defines the line that connects the mean of the EMG datapoints collected during \textit{Rest} to the mean of the projected EMG samples from the $i$th active movement:

\begin{equation}
    L_i = t \left( \mu_i - \mu_{Rest} \right) + \mu_{Rest}
\end{equation}

\noindent where $t \in \left[0, 1\right]$ and $\mu_i$ is the centroid of the feature-extracted EMG datapoints collected during the $i$th movement represented in the LDA subspace. Given an incoming EMG sample $x$, we predicted the movement $\hat{y} \in \left\{ 0, ..., k-1 \right\}$ as the movement whose corresponding axis minimized the distance between $x$ and the orthogonal projection of $x$ onto $L_i$:

\begin{equation}
    \hat{y} = \argmin_i \left\lVert x - \text{proj}_{L_i}x \right\rVert_2 
\end{equation}

\subsection{Training Methods}

All participants were provided with the same baseline PR-based myoelectric system (as described in Section \ref{subsec:patrec}); however, the training method differed between the two participant groups. The experimental group was provided access to the \textit{Reviewer}, a 3D visualization of the pattern recognition decision-space (\textcolor{RoyalBlue}{Fig. \ref{EG_vs_CG}A-C}), while the control group was provided with a virtual environment with a controllable upper limb (\textcolor{RoyalBlue}{Fig. \ref{EG_vs_CG}D-F}).

\subsubsection{Reviewer}
\label{subsubsec:3d_vis}

The \textit{Reviewer} projects the individual's incoming EMG data into the decision-space of the PR-based control method. This is accomplished by first undergoing a standard calibration regimen for PR-based control, where a calibration set of EMG samples is collected to generate a discriminant-based classifier (\textcolor{RoyalBlue}{Fig. \ref{EG_vs_CG}A-C}). Throughout this initialization, features are extracted from windows of raw EMG data, which are then used to generate a low-dimensional basis for classification. During the evaluation periods, subsequent windows of EMG activity are then projected within the same training basis to represent where an individual's current EMG activity lies within the decision-space. Within the visualization, the original training data is represented as colored clusters of data points, and the individual's current EMG activity is represented as a 3D cursor. The cursor's position is modulated in real-time by the user's EMG activity, allowing the participant to directly observe how their changing EMG patterns affect the proximity of their current pattern to the data that the PR was calibrated to. In this way, participants receive direct visual feedback on the discriminability of their calibration data and the repeatability of their control as well as an opportunity to generate and observe how novel patterns of EMG activity map to regions of the decision-space.

\begin{figure*}[!t]
\centerline{\includegraphics[width=0.85\textwidth]{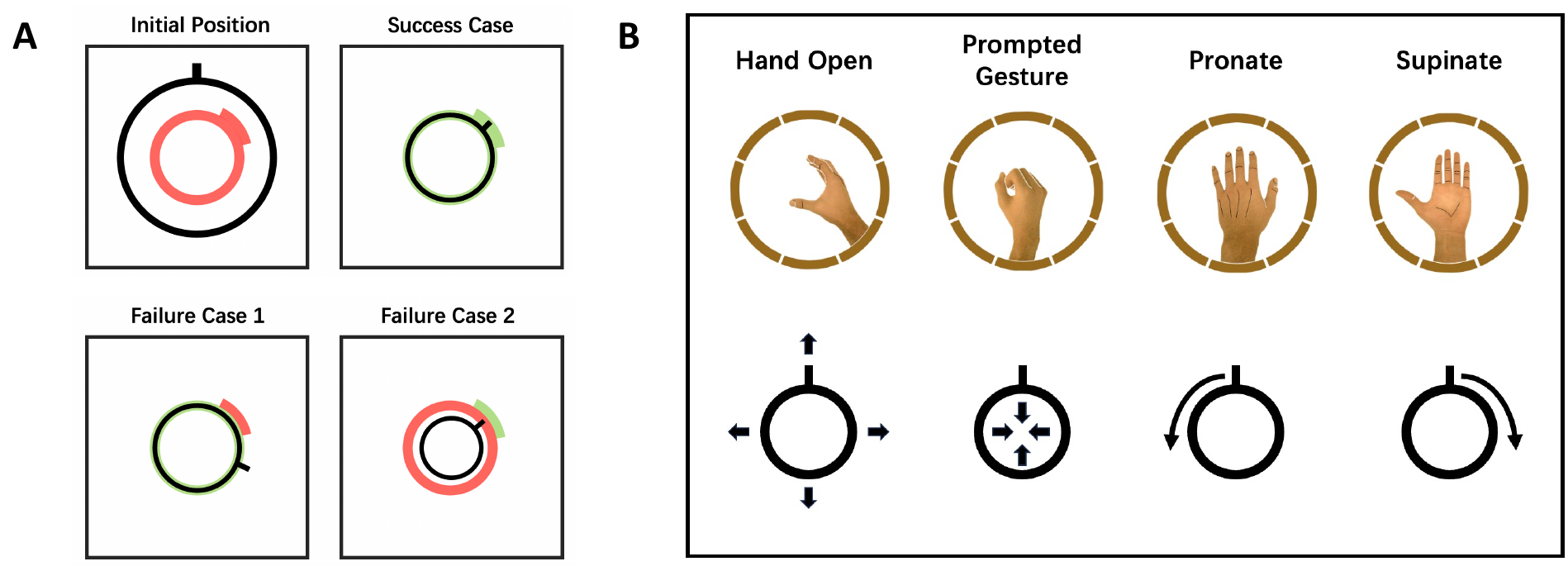}}
\caption{Overview of the Fitts Law Test. (A) Subjects were tasked with aligning a virtual cursor to a target aperture and orientation, displayed as a red ring and protrusion. A trial was deemed successful only when both the ring and protrusion were aligned accurately within a timeframe of 15 seconds. (B) In the FLT, subjects would maneuver a black ring along with a protrusion on the ring. The gestures designated for commanding the ring and protrusion are illustrated above. Upon commencement of a trial, users would be prompted with a gesture to control the ring's closure. The prompted gestures encompass Power Grasp, Key Grasp, Tripod Grasp, Index Point, and Precision Pinch.}
\label{Fittslaw}
\end{figure*}

\subsubsection{Virtual Arm}
\label{subsubsec:virtual_arm}

The virtual arm training method allows participants to operate an anatomical model of the arm as if it were a real-world prosthesis\cite{powell2013user}. Implemented in the Unity game engine (Unity Technologies, San Francisco, CA), the virtual arm receives classification outputs from the PR-based control method and generates 3D animations of the predicted movement. Through this mechanism, participants are provided real-time feedback on the output of their control, analogous to the visual feedback that participants receive when operating a physical upper limb prosthesis (\textcolor{RoyalBlue}{Fig. \ref{EG_vs_CG}D-F}). Participants would rely on the heuristic judgment of the output environment without direct visibility of the decoder's decision boundaries.

\subsection{Experiment Protocols}

To demonstrate the efficacy of the \textit{Reviewer}, we enrolled participants in a 10-session longitudinal study. Additionally, a retention session was held 30 days after the completion of Session 10 to assess the long-term benefits. Participants underwent a PR control training regimen designed to increase in difficulty as it progressed. Each training session progressed through the following general stages: 

\subsubsection{Calibration}

To begin each session, participants collected EMG signals while performing a series of prompted hand and wrist movements. Each movement was trained in a variety of limb positions to reduce limb-position effect in prosthetic control (as dictated by a numbered checkerboard) (\textcolor{RoyalBlue}{Fig. \ref{setup}B})\cite{beaulieu2017multi}. The set of movements trained differed as the protocol progressed. In Sessions 1 – 4, participants were asked to train in five movements: Rest, Hand Open, Power Grasp, Wrist Pronate, and Wrist Supinate. During Sessions 5 – 7, two additional movement classes were added: Tripod Grasp and Key Grasp. Sessions 8 – 10 included the final two movements: Index Point and Precision Pinch.

\subsubsection{Exploration}

After initial calibration, participants were given time to observe their myoelectric control and recalibrate any movements that resulted in misclassification due to overlap between movement groups in the feature space. Participants in the control group were provided visual feedback of their control using the virtual arm (as described in Section \ref{subsubsec:virtual_arm}), while participants in the experimental group were provided visual feedback using the \textit{Reviewer} (as described in Section \ref{subsubsec:3d_vis}). Regardless of the visual feedback presented, participants were allowed to manually recalibrate each separate movement in the Spatial Classifier by adopting new control strategies until satisfied with its performance (or a time limit was reached). Participants were allowed 2 min per calibrated movement (excluding rest) to explore the use of their PR-based control. This resulted in a maximum exploration duration of 8 min for Sessions 1 – 4, 12 min for Sessions 5 – 7, and 16 min for Sessions 8 – 10. At any time, participants could elect to terminate the exploration phase early and progress to the control assessment.

\begin{table*}[b]
\centering
\caption{Fitts' Law Test Results}
\scriptsize  
\setlength{\tabcolsep}{1pt}  
\renewcommand{\arraystretch}{1.4}  
\vspace{2pt}
\begin{tabular}{m{2.7cm}>{\centering\arraybackslash}m{1.45cm} >{\centering\arraybackslash}m{1.45cm} >{\centering\arraybackslash}m{1.45cm} >{\centering\arraybackslash}m{1.45cm} >{\centering\arraybackslash}m{1.45cm} >{\centering\arraybackslash}m{1.45cm} >{\centering\arraybackslash}m{1.45cm} >{\centering\arraybackslash}m{1.45cm} >{\centering\arraybackslash}m{1.45cm} >{\centering\arraybackslash}m{1.45cm}}
\toprule
\rowcolor[HTML]{C0C0C0} 
\textit{\textbf{Metric}} & \textit{\textbf{Session 1}} & \textit{\textbf{Session 2}} & \textit{\textbf{Session 3}} & \textit{\textbf{Session 4}} & \textit{\textbf{Session 5}} & \textit{\textbf{Session 6}} & \textit{\textbf{Session 7}} & \textit{\textbf{Session 8}} & \textit{\textbf{Session 9}} & \textit{\textbf{Session 10}} \\ \midrule

\cellcolor[HTML]{FFFFFF}\multirow{2}{*}{\textbf{\shortstack[l]{Completion Rate (CR)}}} & 
\cellcolor[HTML]{E0E0E0} 0.62 $\pm$ 0.17 & \cellcolor[HTML]{E0E0E0} 0.88 $\pm$ 0.15 & \cellcolor[HTML]{E0E0E0} \textbf{0.94 $\pm$ 0.11} & \cellcolor[HTML]{E0E0E0} 0.97 $\pm$ 0.07 & \cellcolor[HTML]{E0E0E0} \textbf{0.83 $\pm$ 0.10} & \cellcolor[HTML]{E0E0E0} 0.80 $\pm$ 0.07 & \cellcolor[HTML]{E0E0E0} 0.92 $\pm$ 0.12 & \cellcolor[HTML]{E0E0E0} \textbf{0.86 $\pm$ 0.14} & \cellcolor[HTML]{E0E0E0} \textbf{0.89 $\pm$ 0.21} & \cellcolor[HTML]{E0E0E0} \textbf{0.92 $\pm$ 0.07} \\ 
& 0.40 $\pm$ 0.28 & 0.58 $\pm$ 0.31 & \textbf{0.61 $\pm$ 0.27} & 0.78 $\pm$ 0.23 & \textbf{0.52 $\pm$ 0.30} & 0.70 $\pm$ 0.23 & 0.86 $\pm$ 0.07 & \textbf{0.46 $\pm$ 0.31} & \textbf{0.53 $\pm$ 0.27} & \textbf{0.66 $\pm$ 0.25} \\ 
\midrule  

\cellcolor[HTML]{FFFFFF}\multirow{2}{*}{\textbf{\shortstack[l]{Overshoot (OT)}}} & 
\cellcolor[HTML]{E0E0E0} 1.30 $\pm$ 0.65 & \cellcolor[HTML]{E0E0E0} 0.92 $\pm$ 0.58 & \cellcolor[HTML]{E0E0E0} 0.65 $\pm$ 0.44 & \cellcolor[HTML]{E0E0E0} 0.47 $\pm$ 0.59 & \cellcolor[HTML]{E0E0E0} 1.00 $\pm$ 0.52 & \cellcolor[HTML]{E0E0E0} 0.97 $\pm$ 0.19 & \cellcolor[HTML]{E0E0E0} 0.60 $\pm$ 0.41 & \cellcolor[HTML]{E0E0E0} 0.59 $\pm$ 0.17 & \cellcolor[HTML]{E0E0E0} 0.49 $\pm$ 0.21 & \cellcolor[HTML]{E0E0E0} 0.52 $\pm$ 0.18 \\ 
& 1.70 $\pm$ 0.57 & 1.90 $\pm$ 0.57 & 1.50 $\pm$ 0.64 & 1.00 $\pm$ 0.32 & 1.50 $\pm$ 0.54 & 1.40 $\pm$ 0.45 & 0.87 $\pm$ 0.24 & 1.20 $\pm$ 0.31 & 0.83 $\pm$ 0.41 & 0.91 $\pm$ 0.34 \\ 
\midrule

\cellcolor[HTML]{FFFFFF}\multirow{2}{*}{\textbf{\shortstack[l]{Path Efficiency (PE)}}} & 
\cellcolor[HTML]{E0E0E0} 62.17 $\pm$ 7.57 & \cellcolor[HTML]{E0E0E0} 66.00 $\pm$ 6.26 & \cellcolor[HTML]{E0E0E0} 69.50 $\pm$ 6.44 & \cellcolor[HTML]{E0E0E0} 71.83 $\pm$ 7.44 
& \cellcolor[HTML]{E0E0E0} 61.83 $\pm$ 7.17 & \cellcolor[HTML]{E0E0E0} 62.17 $\pm$ 3.37 & \cellcolor[HTML]{E0E0E0} 66.17 $\pm$ 7.60 & \cellcolor[HTML]{E0E0E0} 65.00 $\pm$ 3.74 & \cellcolor[HTML]{E0E0E0} 66.50 $\pm$ 3.45 & \cellcolor[HTML]{E0E0E0} 66.67 $\pm$ 2.42 \\ 
& 56.00 $\pm$ 3.00 & 60.33 $\pm$ 7.06 & 63.00 $\pm$ 7.32 & 63.33 $\pm$ 5.57  
& 57.33 $\pm$ 4.37 & 59.83 $\pm$ 7.33 & 60.00 $\pm$ 4.69 & 62.00 $\pm$ 2.37 & 52.00 $\pm$ 26.37 & 51.17 $\pm$ 26.00 \\ 
\midrule  

\cellcolor[HTML]{FFFFFF}\multirow{2}{*}{\textbf{\shortstack[l]{Throughput (TP)}}} & 
\cellcolor[HTML]{E0E0E0} 0.41 $\pm$ 0.09 & \cellcolor[HTML]{E0E0E0} 0.48 $\pm$ 0.07 
& \cellcolor[HTML]{E0E0E0} 0.48 $\pm$ 0.10 & \cellcolor[HTML]{E0E0E0} 0.55 $\pm$ 0.10 
& \cellcolor[HTML]{E0E0E0} 0.40 $\pm$ 0.05 
& \cellcolor[HTML]{E0E0E0} 0.39 $\pm$ 0.06 & \cellcolor[HTML]{E0E0E0} 0.43 $\pm$ 0.06 & \cellcolor[HTML]{E0E0E0} 0.37 $\pm$ 0.05 & \cellcolor[HTML]{E0E0E0} 0.39 $\pm$ 0.08 & \cellcolor[HTML]{E0E0E0} 0.40 $\pm$ 0.02 \\ 
& 0.28 $\pm$ 0.15 & 0.40 $\pm$ 0.06 
& 0.42 $\pm$ 0.05 & 0.44 $\pm$ 0.05 
& 0.33 $\pm$ 0.04 
& 0.35 $\pm$ 0.06 & 0.38 $\pm$ 0.03 & 0.33 $\pm$ 0.05 & 0.13 $\pm$ 0.55 & 0.11 $\pm$ 0.54 
\\ 
\midrule  

\cellcolor[HTML]{FFFFFF}\multirow{2}{*}{\textbf{\shortstack[l]{Normalized \\ Training Time (NTT)}}} & 
\cellcolor[HTML]{E0E0E0} 0.28 $\pm$ 0.12 & \cellcolor[HTML]{E0E0E0} 0.16 $\pm$ 0.10 & \cellcolor[HTML]{E0E0E0} 0.19 $\pm$ 0.06 & \cellcolor[HTML]{E0E0E0} 0.17 $\pm$ 0.12 & \cellcolor[HTML]{E0E0E0} 0.22 $\pm$ 0.11 & \cellcolor[HTML]{E0E0E0} 0.29 $\pm$ 0.04 
& \cellcolor[HTML]{E0E0E0} 0.20 $\pm$ 0.12 & \cellcolor[HTML]{E0E0E0} 0.21 $\pm$ 0.07 & \cellcolor[HTML]{E0E0E0} 0.16 $\pm$ 0.09 & \cellcolor[HTML]{E0E0E0} 0.19 $\pm$ 0.12 \\ 
& 0.31 $\pm$ 0.10 & 0.19 $\pm$ 0.14 & 0.18 $\pm$ 0.14 & 0.18 $\pm$ 0.13 & 0.16 $\pm$ 0.10 & 0.14 $\pm$ 0.11 
& 0.14 $\pm$ 0.07 & 0.18 $\pm$ 0.06 & 0.12 $\pm$ 0.09 & 0.11 $\pm$ 0.07 \\ 
\midrule  

\cellcolor[HTML]{FFFFFF}\multirow{2}{*}{\textbf{\shortstack[l]{Number of \\ Recalibrations (NR)}}} & 
\cellcolor[HTML]{E0E0E0} 1.70 $\pm$ 2.70 & \cellcolor[HTML]{E0E0E0} 0.67 $\pm$ 1.00 & \cellcolor[HTML]{E0E0E0} 1.30 $\pm$ 2.00 & \cellcolor[HTML]{E0E0E0} 1.30 $\pm$ 1.60 & \cellcolor[HTML]{E0E0E0} 2.70 $\pm$ 1.90 & \cellcolor[HTML]{E0E0E0} 5.50 $\pm$ 3.30 
& \cellcolor[HTML]{E0E0E0} 3.00 $\pm$ 3.10 & \cellcolor[HTML]{E0E0E0} 5.20 $\pm$ 2.90 & \cellcolor[HTML]{E0E0E0} 6.00 $\pm$ 5.10 & \cellcolor[HTML]{E0E0E0} 6.50 $\pm$ 6.00 \\ 
& 0.67 $\pm$ 0.82 & 0.17 $\pm$ 0.41 & 0.17 $\pm$ 0.41 & 0.33 $\pm$ 0.82 & 1.20 $\pm$ 1.50 & 0.67 $\pm$ 1.20 
& 1.30 $\pm$ 0.82 & 2.50 $\pm$ 2.70 & 1.80 $\pm$ 2.40 & 1.30 $\pm$ 1.20 \\ 
\bottomrule
\end{tabular}

\begin{center}
\textbf{Note:}
Completion Rate (CR) are tested at $p < 0.05$ (Student’s t-test), 
and statistically significant differences are shown in \textbf{bold}. 
Secondary metrics (OT, PE, TP, NTT, NR) are evaluated with a post-hoc correction, 
and none met the adjusted significance threshold in this table.
\end{center}

\vspace{8pt}

\begin{tabular}{rl rl}
\fcolorbox{black}{lightgray}{\phantom{000}} & Experimental Group & 
\hspace{20pt} 
\fcolorbox{black}{white}{\phantom{000}} & Control Group  \\
\end{tabular}

\label{tab:numeric_results}
\end{table*}

\subsubsection{Assessment}

Participant myoelectric control proficiency was assessed using a 2D Fitts’ Law Test (FLT) protocol \cite{kamavuako2014usability}. Individuals were centrally positioned, with a computer monitor to their right and a numbered checkerboard to their left (\textcolor{RoyalBlue}{Fig. \ref{setup}A}). The monitor presented each FLT trial, wherein participants were asked to modulate the configuration of a circular cursor using their PR control output. The configuration of the circular cursor is defined as $s=\left( r, \frac{\theta}{2\pi} \right)$ and includes both its normalized aperture ($r \in \left[ 0, 1 \right]$) and angular orientation ($\theta \in \left[ 0, 2\pi \right]$). At the start of a trial, the participant was presented with the circular cursor in an initial configuration, $s_0$, as well as a prompted hand-closing gesture (i.e., Power Grasp, Tripod Grasp, Key Grasp, Index Point, and Precision Pinch). A target configuration was then provided as a red region highlighting the required aperture and orientation (\textcolor{RoyalBlue}{Fig. \ref{Fittslaw}A}). Target configurations were uniformly drawn from the aperture range $\left[ 0.25, 0.75\right]$ and orientation range $\left[ 0.5\pi, 1.5\pi\right]$. Targets were sampled such that: 1) each eligible prosthetic gesture was prompted an equal number of times; and 2) the set of target orientations was balanced between clockwise and counterclockwise rotations.

Given a target configuration, participants were then asked to use their calibrated Spatial Classifier to control the circular cursor. By predicting the prompted hand gesture, the aperture of the cursor would decrease, while predictions of the antagonist movement (i.e., Hand Open) would increase the cursor aperture. Wrist movement predictions (i.e., Wrist Pronate or Wrist Supinate) would rotate the cursor clockwise or counterclockwise, depending on participant handedness (\textcolor{RoyalBlue}{Fig. \ref{Fittslaw}B}).

Starting with the first non-rest movement classification, participants were given 15 seconds to align the circular cursor with the prompted target configuration. A trial was considered successful if the participant was able to output 1 s of continuous RE classifications at the target configuration within the time limit. FLT trials were executed in three locations within the participant's reaching volume: 1) a neutral position directly in front of the participant (i.e., Location 5); 2) reaching towards the periphery at a neutral elevation (i.e., Location 4/6); and 3) reaching across the midline with an elevated arm posture (i.e., Location 3/1; \textcolor{RoyalBlue}{Fig. \ref{setup}B}). The total number of FLT trials during each assessment was determined by the number of calibrated movements: 18 for Sessions 1 – 4 (five calibrated movements), 36 for Sessions 5 – 7 (seven calibrated movements), and 54 for Sessions 8 – 10 (nine calibrated movements).

\subsubsection{Post Washout Retention}

To analyze the long-term impact of the \textit{Reviewer} system, a follow-up evaluation was conducted with participants after a 30-day washout period (i.e., Session 11). During this bonus evaluation, returning participants (5 in the Control Group and 6 in the Experimental Group) were asked to perform the same activities as those performed during Session 10 using the same physical setup.

\subsection{Analysis}

\begin{figure*}[t]
\centerline{\includegraphics[width=0.9\textwidth]{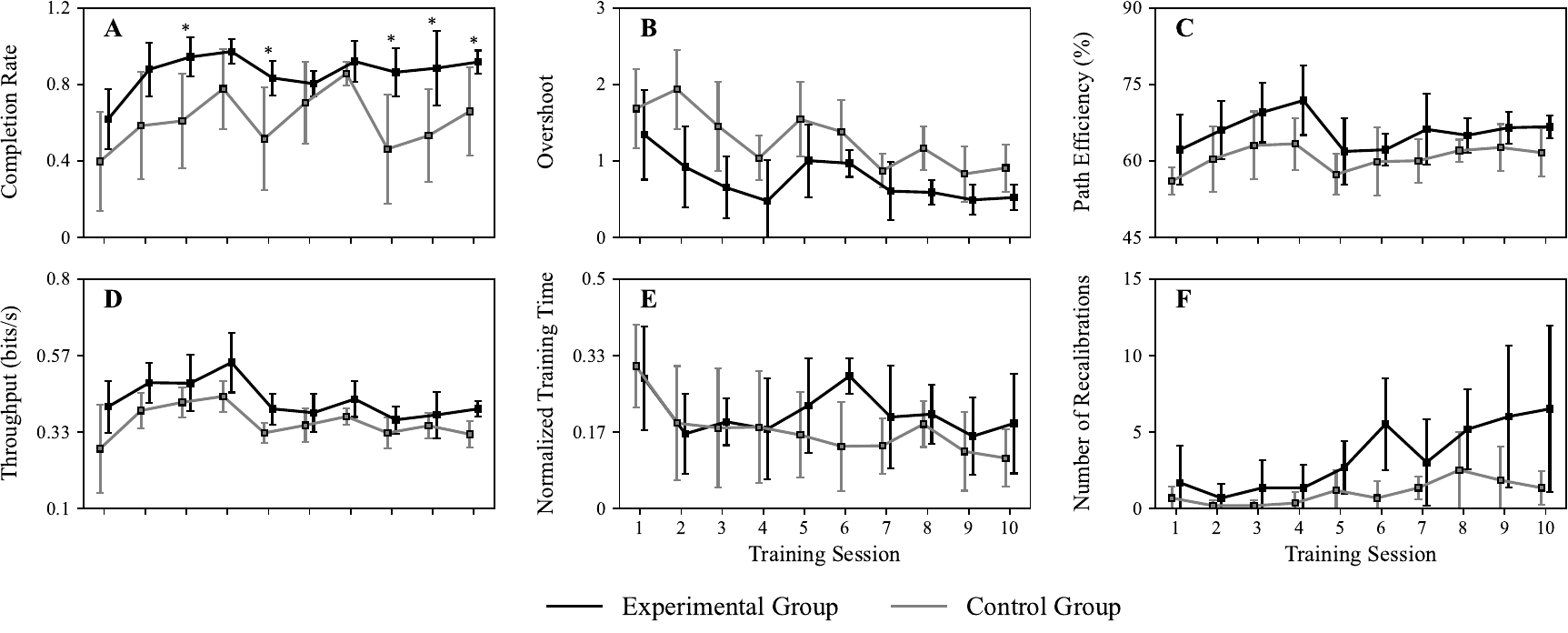}}
\caption{Results of the Experimental and Control Group across the ten-session longitudinal study. In Sessions 1 -- 4, participants were asked to calibrate four active movements (and Rest). In Session 5, two additional movements were incorporated. In Session 8, two further movements were added to increase task difficulty. Throughout each session, the following metrics were computed for each participant: (A) Completion Rate; (B) Overshoot; (C) Path Efficiency; (D) Throughput; (E) Normalized Training Time; and (F) Number of Recalibrations. Asterisks indicate sessions wherein the difference between the Experimental Group and Control Group was significant ($p < 0.05$; Student's t-test).}
\label{results}
\end{figure*}

Control proficiency was evaluated using the following FLT metrics: completion rate (CR), overshoot (OT), path efficiency (PE), and throughput (TP) \cite{scheme2012validation}. CR is defined as the ratio of the number of successfully completed FLT trials ($N_S$) to the number of total FLT trials attempted ($N$):

\begin{equation}
    CR = \frac{N_S}{N}
\end{equation}

\noindent While CR was computed over all $N$ trials, OT, PE, and TP were computed for only the $N_S$ successful trials .

The overshoot count ($N_{OT}$) is defined as the number of moments where the user surpasses the target aperture or orientation. This commonly signifies a shortfall in control precision, an inability to terminate control when appropriate. An overshoot is registered when the user enters the target region (either aperture or orientation) and exits the region in the same direction. OT is reported as the average overshoot count across all successful trials:

\begin{equation}
    OT = \frac{N_{OT}}{N_S}
\end{equation}

PE is the percentage ratio between the optimal path from the cursor's initial ($s_0$) to target ($s_f$) configuration and the actual path traversed by the cursor in configuration-space when manipulated by the participant \cite{williams2008evaluation}. PE describes the participant's economy of control, decreasing with instances of overshoot and movement misclassification. PE is reported as the average path efficiency across all successful trials:

\begin{equation}
    PE = \frac{100}{N_S} \sum_{i=0}^{N_S} \frac{\left\lVert s_{i,0} - s_{i,f} \right\rVert_2}{\sum_{j=0}^{n_i} | s_{i,j} - s_{i,j-1} | }
\end{equation}

\noindent where $s_{i,j}$ represents the $j$th cursor configuration obtained during the $i$th successful FLT trial and $n_i$ represents the number of cursor configurations sampled during execution of the $i$th successful trial.

Throughput acts as a precision-adjusted measure of performance for FLT target acquisition \cite{soukoreff2004towards}. Throughput can be thought of as a task-based measure of user efficiency, where high throughput implies that a user is accomplishing difficult tasks quickly. TP is reported as the average throughput across all successful trials:

\begin{equation}
    TP = \frac{1}{N_S} \sum_{i=1}^{N_S} \frac{1}{T_i} \log_2 \left( \frac{\left\lVert s_{i,0} - s_{i,f} \right\rVert_2}{0.05} + 1\right)
\end{equation}

\noindent where $T_i$ represents the time required to complete the $i$th FLT successful trial.

In addition to the aforementioned FLT metrics, we conducted secondary analyses on the following metrics to quantify subject behavior during the exploration phase of each session: Normalized Training Time (NTT) and Number of Recalibrations (NR).

NTT is defined as the ratio between the actual time the subject spent during the exploration phase ($T_d$) and the maximum allowed exploration duration ($T_{max}$):

\begin{equation}
\text{NTT} = \frac{T_d}{T_{max}}
\end{equation}

We define NR as the total number of recalibrations that occur during the exploration phase of each session. Specifically, any time a participant elects to re-collect data for a particular movement, or replace a previously trained movement, it is counted as one recalibration. Importantly, the time used for these recalibrations is included in the overall duration used to compute NTT. Therefore, participants who choose to recalibrate more frequently do not artificially appear to spend less training time; rather, the additional recalibration effort is reflected in both the NR and NTT metrics.

\section{Results}
\label{sec: results}

Aggregate results of the two participant groups across the 10 protocol sessions are presented in \textcolor{RoyalBlue}{Table \ref{tab:numeric_results}}.

Participants in the Experimental Group accomplished a greater percentage of prompted FLT tasks than those in the Control Group across all protocol sessions, reflected in a greater CR (\textcolor{RoyalBlue}{Fig. \ref{results}A}). We designate CR as primary outcome measures (tested at $\alpha$ = 0.05 without multiple‐comparison adjustment), while OT, PE, TP, NTT, and NR serve as secondary outcomes with post‐hoc corrections. Across Sessions 1 -- 4, when only five movements were eligible for classification (Rest, Hand Open, Power Grasp, Wrist Pronate, and Wrist Supinate), both participant groups exhibited monotonic improvement in CR, although the Experimental Group consistently outperformed the Control Group with a statistically significant difference in Session 3.

In Session 5, when Tripod Grasp and Key Grasp were available to classify, both participant groups experienced a decrease in CR, although the decrease experienced by the Control Group was significantly more severe than that experienced by the Experimental Group. This performance decrease was followed by a period of improvement through Sessions 6 and 7, culminating in the Experimental Group and Control Group having nearly the same CR ($0.92 \pm 0.12$ vs $0.86 \pm 0.07$). However, when Index Point and Precision Pinch were introduced in Session 8, the Control Group again experienced a more significant decrease in CR than that observed with the Experimental Group. While both groups continued to increase their CR across Sessions 8, 9, and 10, the Experimental Group a significant advantage in CR during these evaluations.

\begin{figure}[!t]
\centerline{\includegraphics[width=0.8\columnwidth]{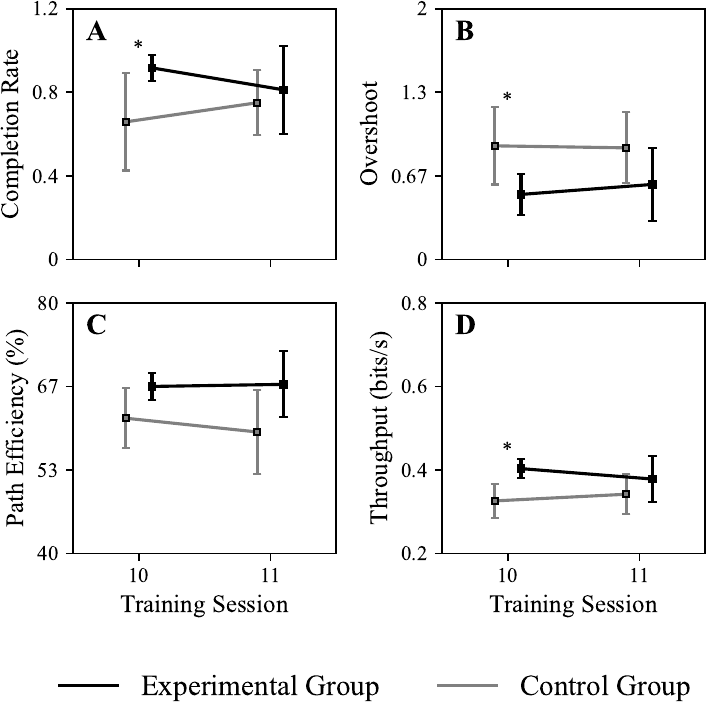}}
\caption{Retention experiment results. Results for (A) completion rate, (B) overshoot, (C) path efficiency, and (D) throughput during Session 11 indicate that participants from both groups performed similarly after the 30-day washout period.}
\label{results_return}
\end{figure}

Furthermore, the Experimental Group consistently completed FLT trials with lower OT relative to the Control Group (\textcolor{RoyalBlue}{Fig. \ref{results}B}). With the addition of new movements in Sessions 5 and 8, both groups exhibited increases in OT; however, the difference in the increase in OT between the two groups was greater in Session 8, when the FLT trials were the most complex (i.e., including the most eligible movements for classification). 

Across the study, the Experimental Group also maintained slightly greater PE when compared to the Control Group with no statistically significant differences after secondary outcome correction, despite a raw difference in Session 4 (\textcolor{RoyalBlue}{Fig. \ref{results}C}). In general, both groups experienced similar performance trends across the ten sessions ($R^2=0.65$). Across Sessions 1 -- 4, there was a monotonic increase in PE with an observable dip in performance occurring in Session 5, when Tripod Grasp and Key Grasp were added to the pool of potential movements. PE began to then increase again across Sessions 5 -- 10. Notably, unlike in Session 5, PE remained relatively stable in Session 8, when two more movements were added (Index Point and Precision Pinch).

Similar to the PE results, the Experimental Group maintained a greater TP when compared to the Control Group (\textcolor{RoyalBlue}{Fig. \ref{results}D}). The trend of the TP results mirror those for PE. The first stage of the protocol (Sessions 1 -- 4) saw a monotonic increase in TP performance, with an observable decrease occurring when new movements were added in Session 5. TP then increased across the second stage of the protocol (Sessions 5 -- 7) with a minor decrease being observed when new movements were again added in Session 8. Across the final stage of the protocol (Sessions 8 -- 10), TP increased (for the Experimental Group) or remained stable (for the Control Group). While the trend of TP results of both groups are highly correlated ($R^2=0.84$), no group differences in TP remained significant once post-hoc corrections were applied, despite raw significance in Sessions 2,4,5, and 10.

Analyses of NTT and NR across the ten sessions revealed complementary aspects of participant behavior during training. During Sessions 1 -- 4, participants in both groups performed similarly, with a general decrease in NTT (\textcolor{RoyalBlue}{Fig. \ref{results}E}) and minimal difference in NR (\textcolor{RoyalBlue}{Fig. \ref{results}F}). During Sessions 5 -- 7, behavior of the Experimental Group began to diverge from the Control Group, with the Experimental Group presenting increases in both NTT and NR but these differences did not survive correction, even though a raw difference was noted in Session 6. During Sessions 8 – 10, the Experimental Group continued to present greater NTT and NR when compared to the Control Group (whose NTT continued to decrease while NR remained low); however, these observations were not significant under the multiple-comparisons framework.

During the optional Session 11, no significant differences were observed between the Experimental and Control Groups in any of the Fitts' Law metrics (\textcolor{RoyalBlue}{Fig. \ref{results_return}}). Additionally, the performance trends seen in Session 10 did not carry over to Session 11. The Experimental Group's performance in nearly all metrics was slightly worse in Session 11 compared to Session 10. In comparison, the Control Group showed modest improvements in CR, OT, and TP, with a slight decrease in PE after the washout period. Eleven of the original twelve participants returned for this session (6 Experimental, 5 Control), with only one participant from the Control Group unable to attend.

\section{Discussion}
\label{sec: discussion}

The discussion will explore the influence of the \textit{Reviewer} system on task performance, the impact of increased control complexity on CR, OT, PE, and TP, and its role in training effectiveness. Unexpected OT findings and signal interference challenges in myoelectric control will be examined. Differences in user adaptation and engagement will be analyzed through NTT and NR. The discussion will also cover the long-term effects of feedback training, study limitations, and future research directions, particularly for individuals with ULL.

\subsection{Control Proficiency}

Throughout all sessions, the means of the reported Fitts' Law metrics (i.e., CR, OT, PE, and TP) indicated that the Experimental Group surpassed the Control Group in all quantified aspects of control proficiency. Because CR serve as our primary metrics, these are interpreted at $\alpha$=0.05 without additional correction. OT, PE, and TP are secondary metrics and thus subject to post-hoc adjustment.  Furthermore, the difference in CR between the two groups increased as the complexity of the FLT increased, indicating that the Experimental Group adapted more swiftly as new movements were made eligible in the control task (i.e., in Sessions 5 and 8). The OT results aligned with this observation, as the Experimental Group experienced a milder increase in OT (when compared to the Control Group) when new movements were added, suggesting better control precision and stability. These findings highlight the effectiveness of the \textit{Reviewer's} visual feedback scheme, facilitating easier adaptation to increasingly complex movement decisions.

Unlike the CR results, for which inter-group differences became more pronounced as more movements were added to the FLT assessments, PE and TP results did not scale with task complexity. Instead, both PE and TP metrics exhibited similar trends between both the Experimental and Control Group, with both groups experiencing similar decreases in PE and TP when new movements were added. In fact, throughout the study, inter-group differences in PE and TP remained nearly constant. This consistent difference between the participant groups suggests that while the Experimental Group exhibited a baseline greater control proficiency than the Control Group (as reflected by the CR results), participant training did not affect the efficiency of control (as measured by PE and TP).

Similarly, OT results were contrary to our expectations. We anticipated that differences in control precision between both groups would increase as more movements were added, which would result in a greater difference in OT as the longitudinal study progressed; however, results instead show a decrease in the inter-group difference in OT. One interpretation of these results is that the increase in FLT complexity instilled a caution in participants that altered their task behavior. During assessments, we observed participants continuously output non-\textit{Rest} control outputs as they modified cursor configurations in simpler FLTs, while for more complex tasks, participants instead opted for intermittent bursts of non-\textit{Rest} control outputs. This strategy adaptation resulted in less OT overall for all participants.

Another interpretation of the OT results is that they are a consequence of the analysis framework, in which OT is only calculated for successful FLTs. Due to the limited trial time (i.e., 15 seconds), participants with low control proficiency were likely to simply fail trials in which overshoot was present, rather than recover and eventually achieve the target. Therefore, for the Control Group, which had lower control proficiency than the Experimental Group, a slight error in control would often result in task failure (resulting in a lower reported OT). This contrasts with the Experimental Group, whose participants were able to successfully complete trials even after producing an initial overshoot due to their greater control proficiency. Under this interpretation, the OT results then highlight a gap in the analysis framework regarding participant behavior during task failure (see \textcolor{RoyalBlue}{Subsection \ref{subsec:limits}}).

NTT and NR (also secondary metrics) did not remain significant under multiple-comparisons correction, but the secondary analysis showed that the Experimental Group spent more time in the exploration phase and updated their movements more frequently than the Control Group. From anecdotal feedback, participants in the Experimental Group attributed their increased recalibration frequency to the 3D visualization, which made misclassifications more apparent and encouraged them to refine problematic gestures. Thus, higher NTT and NR appear to be user-driven reactions to clearer feedback, rather than mere random increases in recalibration. Post-study interviews also revealed that the Control Group found recalibration futile during later sessions due to the lack of specific performance feedback.  In contrast, the \textit{Reviewer} provided the Experimental Group with clear feedback on control proficiency, incentivizing improvement by allowing users to observe tangible distinctions in class separability during movement updates. These findings highlight the effectiveness of the \textit{Reviewer} system in encouraging continuous engagement and iterative improvements in control proficiency.

Based on anecdotal feedback from participants, the Experimental Group reported less difficulty in modifying and fine-tuning movements, attributed to the spatial visualization of PR results. This visualization enabled them to perceive the impact of modulating their muscle contractions on the Spatial Classifier's decision-space, fostering a more informed recalibration process. Using the \textit{Reviewer}, participants were also able to refine their control strategies effectively by observing movement overlaps that resulted in misclassifications. In contrast, the Control Group encountered confusion while adding or updating control movements and had a less robust understanding of which specific movements were causing confusion. In certain instances, their efforts to update gestures inadvertently deteriorated their performance by introducing conflicts with other movements that were not initially problematic. The \textit{Reviewer} system granted the subjects a clear map between their muscle contractions and its resultant position within the decision-space and offered a platform to explore and identify the most optimal replacement gesture before making changes. The \textit{Reviewer} demonstrates its potential in aiding individuals with ULL by addressing this common challenges associated with myoelectric control.

\subsection{Long-Term Effects of Feedback Training}
 
 The slight performance decline in the Experimental Group from Session 10 to Session 11 was anticipated, as it is consistent with the literature on adaptation and skill retention. On the other hand, when questioned about post-washout performance, several Control Group participants expressed that they did not remember the strategies previously employed during Sessions 1 -- 10 and used Session 11 as an opportunity to try completely novel strategies for pattern generation. Such a reevaluation explains the modest improvements observed by the Control Group as a whole. In either case, results of the post-washout assessment suggest that the benefits gained using either visualization paradigm do not persist without habitual training, a notion supported in the literature \cite{resnik2018evaluation, dyson2020learning}.

\subsection{Study Limitations and Future Directions}
\label{subsec:limits}

This study was not without its limitations. One significant limitation is that many of the Fitts' Law metrics were only calculated for successful trials, thus omitting detailed performance data from unsuccessful trials. This approach means that valuable insights into user behavior and system performance during task failures are not captured in our current analysis framework. For some metrics (e.g., OT), it is simple to extend analysis to unsuccessful trials; however, for others (e.g., PE and TP), their application to failed trial data are not useful by definition. In future research, we plan to address this gap in analysis by considering metrics designed to provide insight on task failure under the FLT paradigm (e.g.,  time to first error, mean movement speed during failure) \cite{wobbrock2008error}.

Relatedly, another limitation of this study is the absence of a metric that captures failures to initialize an intended movement. Often, users may successfully and accurately complete an FLT trial, but only after several attempts to produce the intended movement. Introducing a metric that evaluates the percentage of correct movements performed in each independent trial could provide valuable insights. Such a metric would be crucial for assessing the accuracy of control and gauging user proficiency with the system. However, implementing this metric presents challenges. The FLT involves two degrees of freedom — aperture and orientation — making it difficult to automatically determine the subject's true intentions. Difficulty is compounded by the allowance of target overshoot, wherein the optimal control output then switches to undo the previous cursor configuration changes. A promising metric to capture the complexities of FLT behavior is perhaps \textit{effective movement percentage}, or the percentage of control outputs that moved the cursor closer to the target configuration during a FLT.

Since this study was performed on individuals with intact limbs, further studies involving individuals with ULL are essential to determine if the patterns observed in individuals with intact limbs can be effectively transferred to individuals with ULL. The unique musculature, sensory, and proprioceptive feedback inherent to individuals with ULL could potentially lead to non-generalizable results\cite{lotze2001phantom, ostlie2011assessing}. In particular, proprioceptive feedback plays a crucial role in the context of this study, especially in task-driven tests where specific gestures require execution with a prescribed level of proportionality\cite{windhorst2007muscle}.

\begin{figure*}[t]
\centerline{\includegraphics[width=0.9\textwidth]{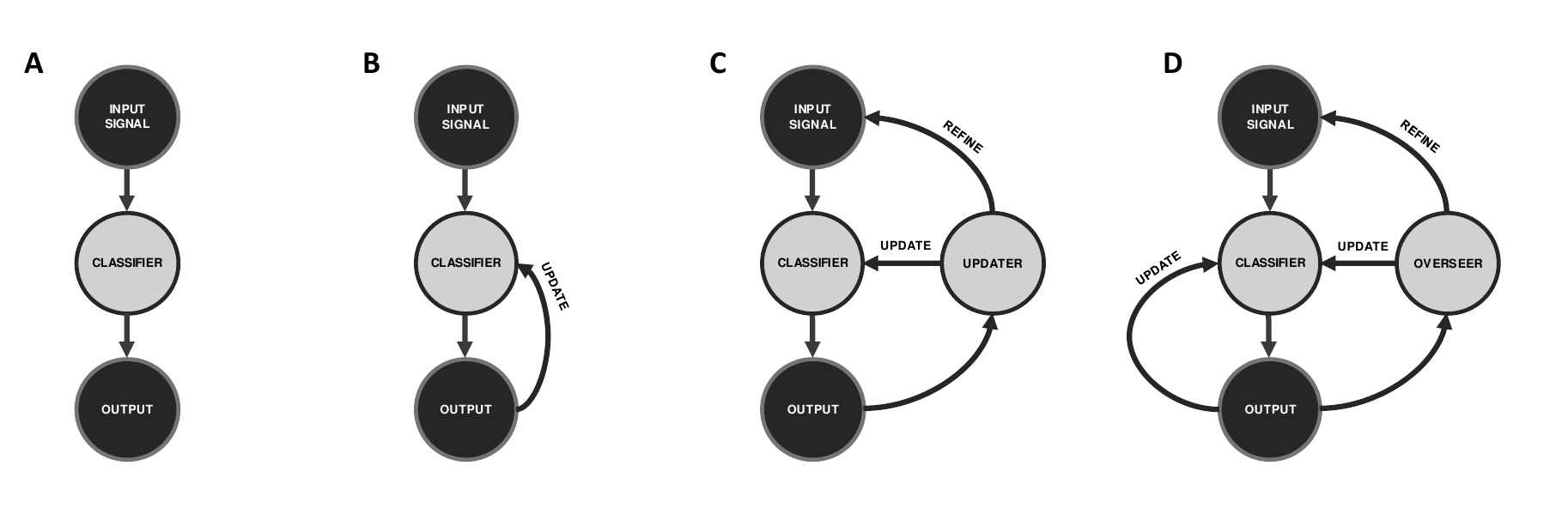}}
\caption{ PR Control Architectures. (A) Traditional PR Decoding: Input signals pass through a static classifier to generate output, requiring users to adapt entirely to the classifier.
(B) Adaptive PR Decoding: The classifier is dynamically updated according to the effectiveness of its output.
(C) Human-in-the-loop PR Training and Decoding: Input EMG signals are classified as before, but now real-time visual feedback of the decision space allows users to directly observe the effect of their EMG signals on classifier processing. Users can then iteratively refine their control through targeted adjustments to both their input signals and the classifier's decision space.
(D) Proposed Future Directions: Ideally, the prosthesis user reduces direct intervention by relying on a lower level autonomous subunit that self-adapts the classifier based on the effectiveness of its output. The user then evolves into an overseer role, intervening minimally yet effectively to improve decoder performance.}
\label{rev}
\end{figure*}

Based on this study's findings, we propose a generalized overview of PR training frameworks to suggest a potential PR co-adaptive training architecture. Traditional PR systems rely on static decoders to interpret input signals (\textcolor{RoyalBlue}{Fig. \ref{rev}A}), but these rigid frameworks prevent any real-time adjustments. In response, traditional work in this field has explored adaptive strategies that dynamically modify the classifier based on classifier output \cite{vidovic2015improving, amsuss2013self, sensinger2009adaptive} (\textcolor{RoyalBlue}{Fig. \ref{rev}B}). However, these approaches largely exclude the user from the adaptation loop, making it difficult to correct certain errors and limiting clinical feasibility.

In contrast, the present study introduces a human-in-the-loop co-adaptation framework (\textcolor{RoyalBlue}{Fig. \ref{rev}C}), wherein the user can observe the decoding space to continuously modify both the input signals and the classifier directly to iteratively improve control. While this approach significantly enhances control proficiency over time, it ultimately requires the user to spend increasing effort on recalibration as task complexity grows, thereby diminishing gains.

Thus, we suggest that future work should focus on co-adaptive strategies that minimize constant user recalibration. Ideally, the system would support two levels of adaptation. An autonomous self-adaptation subunit can be responsible for high-frequency modifications in response to real-time classifier performance, while the prosthesis user will be responsible for low-frequency updates that are sparse but more substantial. Such an approach could increase adaptation speed and accuracy, enabling the system to handle increasing control complexity without imposing a large burden on the user (\textcolor{RoyalBlue}{Fig. \ref{rev}D}).

\section{Conclusion}

The results of this study demonstrate that the use of 3D visual feedback that is tied to the intrinsic decision-space of a pattern recognition-based myoelectric decoder (e.g., the \textit{Reviewer} system) outperforms the use of standard clinical visualization methods (e.g., virtual arm) in increasing the control proficiency of na{\"i}ve prosthesis users. In particular, results suggest the following benefits:

\begin{enumerate}
  \item improved functionality, as quantified by an increased completion rate (as well as improvements in other secondary metrics) across a series of Fitts' Law Test assessments;
  \item improved control scalability, as quantified by an attenuated decrease in completion rate when new movements are introduced (i.e., Sessions 5 and 8);
  \item improved user engagement and motivation during training, as quantified by an increased number of recalibrations in preparation for the most complex control assessments (i.e., Sessions 8 -- 10);
\end{enumerate}

\noindent  By offering intuitive, real-time feedback within the decoder’s decision-space, the proposed visualization approach reduces cognitive load and minimizes user reliance on subjective guesswork. Moreover, enabling dynamic, data-driven recalibration introduces a more adaptive and self-correcting training framework, leading toward an increasingly automated interaction paradigm. Collectively, this paradigm advances PR-controlled prostheses toward greater accessibility, user-friendliness, and long-term functional integration.

\section*{Acknowledgment}
The authors extend their sincere gratitude to the human subjects who participated in this study. This research was partially supported by a grant from the National Institutes of Health (NIH) awarded jointly to Johns Hopkins University (JHU) and Infinite Biomedical Technologies (IBT). Co-author Thakor is a co-founder of IBT; his affiliation and potential conflict of interest have been disclosed to the relevant institutions, with conflict management overseen by JHU.




\begin{thebibliography}{40}

\bibitem{ziegler2008estimating}
K.\~Ziegler-Graham, E.\~J. MacKenzie, P.\~L. Ephraim, T.\~G. Travison, and R.\~Brookmeyer, “Estimating the prevalence of limb loss in the United States: 2005 to 2050,” \emph{Arch. Phys. Med. Rehabil.}, vol.\~89, no.\~3, pp. 422–429, 2008.

\bibitem{avalere2024}
M.~Caruso and S.~Harrington, “Prevalence of limb loss and limb difference in the United States: Implications for public policy,” \emph{Avalere}, Feb. 2024. [Online]. Available: \url{https://avalere.com/wp-content/uploads/2024/02/Prevalence-of-Limb-Loss-and-Limb-Difference-in-the-United-States_Implications-for-Public-Policy.pdf}

\bibitem{darter2018factors}
B.~J. Darter, C.~E. Hawley, A.~J. Armstrong, L.~Avellone, and P.~Wehman, “Factors influencing functional outcomes and return-to-work after amputation: A review,” \emph{J. Occup. Rehabil.}, vol.~28,
pp. 656–665, 2018.

\bibitem{geethanjali2016myoelectric}
P.\~Geethanjali, “Myoelectric control of prosthetic hands: State-of-the-art review,” \emph{Med. Devices: Evidence and Research}, pp. 247–255, 2016.

\bibitem{parajuli2019real}
N.~Parajuli \emph{et al.}, “Real-time EMG-based pattern-recognition
control for hand prostheses: A review,” \emph{Sensors}, vol.~19, no.~20,
p.~4596, 2019.

\bibitem{toledo2019support}
D.~C. Toledo-Pérez, J.~Rodríguez-Reséndiz, R.~A. Gómez-Loenzo, and
J.~C. Jauregui-Correa, “Support-vector-machine-based EMG-signal
classification techniques: A review,” \emph{Appl. Sci.}, vol.~9, no.~20, p.~4402, 2019.

\bibitem{alkan2012identification}
A.~Alkan and M.~Günay, “Identification of EMG signals using discriminant analysis and SVM classifier,” \emph{Expert Syst. Appl.}, vol.~39, no.~1, pp. 44–47, 2012.

\bibitem{hudgins1993new}
B.~Hudgins, P.~Parker, and R.~N. Scott, “A new strategy for multifunction myoelectric control,” \emph{IEEE Trans. Biomed. Eng.}, vol.~40, no.~1, pp. 82–94, 1993.

\bibitem{phinyomark2012feature}
A.~Phinyomark, P.~Phukpattaranont, and C.~Limsakul, “Feature reduction and selection for EMG-signal classification,” \emph{Expert Syst. Appl.}, vol.~39, no.~8, pp. 7420–7431, 2012.

\bibitem{kuiken2016comparison}
T.~A. Kuiken, L.~A. Miller, K.~Turner, and L.~J. Hargrove, “A comparison of pattern-recognition and direct control of a multi-degree-of-freedom transradial prosthesis,” \emph{IEEE J. Transl. Eng. Health Med.}, vol.~4, pp. 1–8, 2016.

\bibitem{bouwsema2014changes}
H.~Bouwsema, C.~K. van der Sluis, and R.~M. Bongers, “Changes in performance over time while learning to use a myoelectric prosthesis,”
\emph{J. Neuroeng. Rehabil.}, vol.~11, pp. 1–15, 2014.

\bibitem{scheme2011electromyogram}
E.~Scheme and K.~Englehart, “Electromyogram pattern recognition for powered upper-limb prostheses: State of the art and challenges,”
\emph{J. Rehabil. Res. Dev.}, vol.~48, no.~6, 2011.

\bibitem{bunderson2012quantification}
N.~E. Bunderson and T.~A. Kuiken, “Quantification of feature-space changes with experience during EMG pattern-recognition control,”
\emph{IEEE Trans. Neural Syst. Rehabil. Eng.}, vol.~20, no.~3,
pp. 239–246, 2012.

\bibitem{kristoffersen2019effect}
M.~B. Kristoffersen, A.~W. Franzke, C.~K. van der Sluis, A.~Murgia, and
R.~M. Bongers, “Effect of feedback during training on pattern-recognition prosthesis control,” \emph{IEEE Trans. Neural Syst. Rehabil. Eng.}, vol.~27, no.~10, pp. 2087–2096, 2019.

\bibitem{powell2013training}
M.~A. Powell and N.~V. Thakor, “A training strategy for learning
pattern-recognition control for myoelectric prostheses,” \emph{JPO},
vol.~25, no.~1, pp. 30–41, 2013.

\bibitem{he2015user}
J.~He \emph{et al.}, “User adaptation in long-term, open-loop
myoelectric training,” \emph{J. Neural Eng.}, vol.~12, no.~4,
p. 046005, 2015.

\bibitem{powell2013user}
M.~A. Powell, R.~R. Kaliki, and N.~V. Thakor, “User training for
pattern-recognition myoelectric prostheses,” \emph{IEEE Trans.
Neural Syst. Rehabil. Eng.}, vol.~22, no.~3, pp. 522–532, 2013.

\bibitem{dyson2020learning}
M.~Dyson, S.~Dupan, H.~Jones, and K.~Nazarpour, “Learning and scalability of abstract myoelectric control,” \emph{IEEE Trans. Neural Syst. Rehabil. Eng.}, vol.~28, no.~7, pp. 1539–1547, 2020.

\bibitem{davoodi2011real}
R.~Davoodi and G.~E. Loeb, “Real-time animation software for customized
training to use motor-prosthetic systems,” \emph{IEEE Trans. Neural Syst. Rehabil. Eng.}, vol.~20, no.~2, pp. 134–142, 2011.

\bibitem{van2016learning}
L.~Van Dijk, C.~K. van der Sluis, H.~W. Van Dijk, and R.~M. Bongers,
“Learning an EMG-controlled game: Task-specific adaptations and transfer,” \emph{PLoS One}, vol.~11, no.~8, p. e0160817, 2016.

\bibitem{winslow2018mobile}
B.~D. Winslow, M.~Ruble, and Z.~Huber, “Mobile, game-based training for
myoelectric prosthesis control,” \emph{Front. Bioeng. Biotechnol.}, vol.~6, p.~94, 2018.

\bibitem{Ottobock}
Ottobock, “Myo Plus app,” 2020. [Online]. Available: \url{https://www.ottobock.com/en/apps/myoplusapp/myo-plus-app-de.html}

\bibitem{de2020guiding}
E.~de Montalivet, K.~Bailly, A.~Touillet, N.~Martinet, J.~Paysant, and
N.~Jarrasse, “Guiding the training of users with pattern-similarity
biofeedback,” \emph{IEEE Trans. Neural Syst. Rehabil. Eng.}, vol.~28, no.~8, pp. 1731–1741, 2020.

\bibitem{nawfel2022influence}
J.~L. Nawfel, K.~B. Englehart, and E.~J. Scheme, “Influence of training with visual biofeedback on myoelectric usability,” \emph{IEEE Trans. Neural Syst. Rehabil. Eng.}, vol.~30, pp. 878–892, 2022.

\bibitem{levay2024pattern}
G.~M. Levay \emph{et al.}, “Pattern separability visual feedback to improve pattern-recognition decoding performance,” in \emph{Proc. MCU Myoelectric Controls Symp.}, 2024.

\bibitem{englehart1999classification}
K.~Englehart, B.~Hudgins, P.~A. Parker, and M.~Stevenson, “Classification of the myoelectric signal using time-frequency representations,” \emph{Med. Eng. Phys.}, vol.~21, no.~6–7, pp. 431–438, 1999.

\bibitem{englehart2003robust}
K.~Englehart and B.~Hudgins, “A robust, real-time control scheme for
multifunction myoelectric control,” \emph{IEEE Trans. Biomed. Eng.},
vol.~50, no.~7, pp. 848–854, 2003.

\bibitem{beaulieu2017multi}
R.~J. Beaulieu \emph{et al.}, “Multi-position training improves robustness
of pattern recognition and reduces limb-position effect,” \emph{J.
Prosthet. Orthot.}, vol.~29, no.~2, pp. 54–62, 2017.

\bibitem{kamavuako2014usability}
E.~N. Kamavuako, E.~J. Scheme, and K.~B. Englehart, “On the usability of
intramuscular EMG for prosthetic control: A Fitts’ law approach,”
\emph{J. Electromyogr. Kinesiol.}, vol.~24, no.~5, pp. 770–777, 2014.

\bibitem{scheme2012validation}
E.~J. Scheme and K.~B. Englehart, “Validation of a selective
ensemble-based classification scheme,” \emph{IEEE Trans.
Neural Syst. Rehabil. Eng.}, vol.~21, no.~4, pp. 616–623, 2012.

\bibitem{williams2008evaluation}
M.~R. Williams and R.~F. Kirsch, “Evaluation of head orientation and neck
muscle EMG as command inputs for individuals with tetraplegia,”
\emph{IEEE Trans. Neural Syst. Rehabil. Eng.}, vol.~16, no.~5,
pp. 485–496, 2008.

\bibitem{soukoreff2004towards}
R.~W. Soukoreff and I.~S. MacKenzie, “Towards a standard for pointing-device
evaluation,” \emph{Int. J. Hum.-Comput. Stud.}, vol.~61, no.~6,
pp. 751–789, 2004.

\bibitem{resnik2018evaluation}
L.~Resnik \emph{et al.}, “Evaluation of EMG pattern recognition for upper-limb prosthesis control,” \emph{J. Neuroeng. Rehabil.}, vol.~15,
pp. 1–13, 2018.

\bibitem{wobbrock2008error}
J.~O. Wobbrock, E.~Cutrell, S.~Harada, and I.~S. MacKenzie, “An error model for pointing based on Fitts’ law,” in \emph{Proc. CHI Conf.
Human Factors Comput. Syst.}, 2008, pp. 1613–1622.

\bibitem{lotze2001phantom}
M.~Lotze, H.~Flor, W.~Grodd, W.~Larbig, and N.~Birbaumer, “Phantom movements and pain: An fMRI study,” \emph{Brain}, vol.~124, no.~11, pp. 2268–2277, 2001.

\bibitem{ostlie2011assessing}
K.~Østlie \emph{et al.}, “Assessing physical function in adult acquired
major upper-limb amputees,” \emph{Arch. Phys. Med. Rehabil.},
vol.~92, no.~10, pp. 1636–1645, 2011.

\bibitem{windhorst2007muscle}
U.~Windhorst, “Muscle proprioceptive feedback and spinal networks,”
\emph{Brain Res. Bull.}, vol.~73, no.~4–6, pp. 155–202, 2007.

\bibitem{vidovic2015improving}
M. M.-C. Vidovic \emph{et al.}, “Improving the robustness of myoelectric
pattern recognition by covariate-shift adaptation,” \emph{IEEE Trans.
Neural Syst. Rehabil. Eng.}, vol.~24, no.~9, pp. 961–970, 2016.

\bibitem{amsuss2013self}
S.~Amsüss \emph{et al.}, “Self-correcting pattern-recognition system of
surface EMG for upper-limb prosthesis control,” \emph{IEEE Trans.
Biomed. Eng.}, vol.~61, no.~4, pp. 1167–1176, 2014.

\bibitem{sensinger2009adaptive}
J.~W. Sensinger, B.~A. Lock, and T.~A. Kuiken, “Adaptive pattern recognition
of myoelectric signals,” \emph{IEEE Trans. Neural Syst.
Rehabil. Eng.}, vol.~17, no.~3, pp. 270–278, 2009.

\end{thebibliography}
\end{document}